\pdfoutput=1
\documentclass[aps,twocolumn, floatfix, prl]{revtex4}
\usepackage{graphicx}
\DeclareGraphicsExtensions{.pdf}
\usepackage{amsmath,amssymb,bbold,bm,color}
\usepackage{float}
\usepackage{epstopdf}

\newcommand{\bk}{{\bm k}}

\newcommand{\bQ}{{\bm Q}}

\newcommand{\bv}{{\bm v}}
\newcommand{\bu}{{\bm u}}

\newcommand{\bs}{{\bm s}}

\newcommand{\cT}{{\cal T}}
\newcommand{\cK}{{\cal K}}

\newcommand{\cP}{{\cal P}}
\newcommand{\cH}{{\cal H}}

\newcommand{\cN}{{\cal N}}

\newcommand{\bee}{\begin{equation}}
\newcommand{\ee}{\end{equation}}

\begin{document}

\title{Superconducting proximity effect and Majorana flat bands at the surface of a Weyl semimetal}

\author{Anffany Chen}
\author{M. Franz}
\affiliation{Department of Physics and Astronomy, University of
British Columbia, Vancouver, BC, Canada V6T 1Z1}
\affiliation{Quantum Matter Institute, University of British Columbia, Vancouver BC, Canada V6T 1Z4}

\begin{abstract} 
We study the proximity effect between an $s$-wave superconductor (SC) and the surface states of a Weyl semimetal. An interesting two-dimensional SC forms in such an interface with properties resembling in certain aspects the Fu-Kane superconductor 
with some notable differences. In a Weyl semimetal with unbroken time reversal symmetry the interface SC supports completely flat Majorana bands in a linear Josephson junction with a $\pi$ phase difference.  We discuss stability of these bands against disorder and propose ways in which they can be observed experimentally. 
\end{abstract}

\date{\today}

\maketitle
Interfacing topological materials with conventionally ordered states of matter, such as magnets and superconductors, has led to important conceptual advances over the past decade. Notable examples of this approach include the Fu-Kane superconductor \cite{fu1}  that occurs in the interface of a 3D strong topological insulator (STI) and a conventional $s$-wave SC,
 the ``fractional'' quantum Hall effect that arises when STI is interfaced with a magnetic insulator \cite{fu2,essin1,qi2}, as well as many interesting phenomena that occur when both SC and magnetic domains are present \cite{akhmerov1,fu4}. Rich physics, including Majorana zero modes, also results when the edge of a 2D topological insulator is interfaced with magnets and superconductors \cite{nilsson1,fu3,mi1}. More recently various exotic phases of quantum matter have been predicted to emerge based on these same ingredients in strongly interacting systems \cite{clarke1,lindner1,vaezi1,barkeshli1,wang1,metlitski1,bonderson1,chen1,mong1,milsted1,rahmani1}.

In this Letter we explore the physics of the interface between a Weyl semimetal and an $s$-wave superconductor. Surface states of a Weyl semimetal exhibit characteristic Fermi arcs that terminate at surface projections of the bulk Weyl nodes \cite{wan1,vafek1}. Such ``open'' Fermi surfaces are fundamentally impossible in a purely 2D system and we thus expect the resulting SC state to also be anomalous. In this respect the situation is similar to the Fu-Kane superconductor \cite{fu1} whose existence hinges on the odd number of Dirac fermions present on the surface of an STI. As we shall see there are several notable differences between STI/SC and Weyl/SC interfaces which make the latter a distinct and potentially more versatile platform for explorations of new phenomena.

\begin{figure}[t]
\includegraphics[width = 8.5cm]{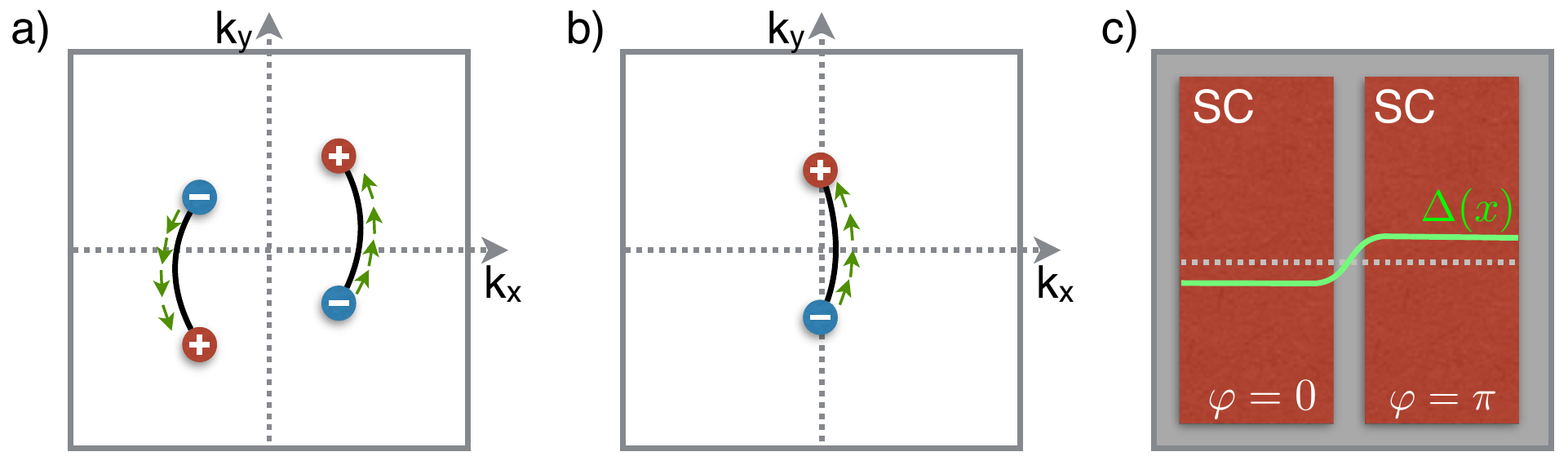}
\caption{Surface Fermi arcs in a minimal model of a Weyl semimetal with (a) and without (b) time reversal symmetry $\cT$. Weyl nodes are represented by circles with positive (negative) chirality; green arrows indicate the possible direction of electron spin along the arcs. Panel (c) shows the $\pi$ junction setup on the surface of a Weyl semimetal.
}\label{fig1}
\end{figure}
Nondegenerate Weyl points can occur in crystals with broken time reversal symmetry $\cT$ or broken bulk inversion symmetry $\cP$. Recent experimental work reported convincing evidence for Weyl nodes and surface Fermi arcs in $\cT$-preserving noncentrosymmetric crystals in the TaAs family of semimetals \cite{hasan1,lv1,yan1,lv2,yang1,hasan2}. We therefore focus our discussion on the SC proximity effect in this class of materials. When a crystal respects $\cT$ the minimum number of Weyl points $\cN$ is 4. This is because under $\cT$ a Weyl point at crystal momentum $\bQ$ maps onto a Weyl point at $-\bQ$ with the same chirality. Since the total chiral charge in the Brillouin zone must vanish there has to be another pair of $\cT$-conjugate Weyl nodes with an opposite chirality. We begin by discussing the SC proximity effect in this minimal case with $\cN=4$. We note that although the currently known Weyl semimetals exhibit larger number of nodes ($\cN=24$ in the TaAs family) recent theoretical work identified materials that could realize instances with smaller $\cN$ \cite{fang1,ruan1}, including the minimal case with $\cN=4$ predicted in MoTe$_2$ \cite{bernevig1}.  The generic surface state of such a minimal $\cT$-preserving Weyl semimetal is depicted in Fig.\ \ref{fig1}a. 

Interfacing such a surface with a spin singlet $s$-wave SC one expects formation of a paired state from time-reversed Bloch electrons at crystal momenta $\bk$ and $-\bk$ along the arcs. A minimal model describing this situation is defined by the second quantized Bogoliubov-de Gennes (BdG) Hamiltonian $\cH=\sum_\bk\hat\Psi^\dagger_\bk h_{\rm BdG}(\bk)\hat\Psi_\bk$ where
$\hat\Psi_\bk=(c_{\bk\uparrow},c_{\bk\downarrow},c^\dagger_{-\bk\downarrow},-c^\dagger_{-\bk\uparrow})^T$
is the Nambu spinor and   
\begin{equation}\label{h1}
h_{\rm BdG}= \begin{pmatrix}
h_0(\bk)-\mu & \Delta \\
\Delta^\dagger &-s^y h_0^*(-\bk)s^y+\mu
\end{pmatrix},
\end{equation}
and $\mu$ is the chemical potential. 
The normal state surface Hamiltonian for two parallel Fermi arcs with electron spin locked perpendicular to its momentum as in Fig.\ \ref{fig1}a  can be written as 
\begin{equation}\label{h2}
h_0(\bk)=vs^zk_x-\eta, \ \ \ \ {\rm for} \ |k_y|<K. 
\end{equation}
Here $s^\alpha$ are Pauli matrices acting in electron spin space. One can easily check that Eq.\ (\ref{h2}) produces two parallel Fermi arcs of length $2K$ along $k_y$ separated by distance $\eta/v$ in the $k_x$ direction when $\mu=0$.
It is possible to make this model more generic by allowing the velocity $v$ and parameter $\eta$ to depend on $k_y$ and the cutoff $K$ on $k_x$ (which would curve and shift the arcs similar to Fig.\ \ref{fig1}a) but the minimal model defined above already captures the essential physics we wish to describe. The surface model (\ref{h2}) does not capture the bulk bands that become gapless near the Weyl points.  The model based on Eq.\ (\ref{h1}) is therefore valid only away from the surface projections of the Weyl points, specifically when $v||k_y|-K|>\Delta$, and in the absence of large momentum transfer disorder scattering that would mix the surface and bulk low-energy states. 

If we define another set of Pauli matrices $\tau^\alpha$ in the Nambu space we can write Eq.\ (\ref{h1}) for each  $|k_y|<K$ as
\begin{equation}\label{h3}
h_{\rm BdG}=(vs^zk_x-\tilde\mu)\tau^z+\Delta_1\tau^x-\Delta_2\tau^y,
\end{equation}
where $\tilde\mu=\mu+\eta$ and $\Delta=\Delta_1+i\Delta_2$. For spatially uniform $\Delta$ the spectrum of  $h_{\rm BdG} $ is fully gapped. Its topological character can be exposed by noting that  for each allowed value of $k_y$ Eq.\ (\ref{h3}) coincides with the Hamiltonian describing the edge of a 2D TI in contact with a SC. It is well known that unpaired Majorana zero modes (MZMs) exist in such an edge at domain walls between SC and magnetic regions \cite{fu3}. This is because $\cT$-breaking also opens up a gap in 2D TI edge and MZMs occur at  boundaries between two differently gapped regions.  In our present case of the Weyl semimetal breaking $\cT$ generically does not open a gap. However, we can achieve a similar result by creating a domain wall in the complex order parameter $\Delta(x)$. Specifically, we show below  that a $\pi$ junction (i.e.\ a boundary between two regions whose SC phase $\varphi$ differs by $\pi$, see Fig.\ \ref{fig1}c) hosts a {\em pair} of protected MZMs (one for each arc and for each allowed value of $k_y$), separated by a gap from the rest of the spectrum. As a function of $k_y$, then, the band structure of such a $\pi$ junction exhibits a flat Majorana band pinned to zero energy. We argue below that such flat Majorana bands are experimentally observable and represent a generic robust property of any $\cT$-preserving  Weyl/SC interface. 

To exemplify this property within the effective surface theory (\ref{h3}) we replace $k_x\to -i\partial_x$ and assume $\Delta(x)$ to be purely real with a soliton profile such that $\Delta(x)\to\pm\Delta_0$ as $x\to\pm\infty$.  We next note that Hamiltonian (\ref{h3}) can be brought to a block diagonal form $h_{\rm BdG}= {\rm diag}(h_+,h_-)$ by a unitary transformation that exchanges its second and third rows and columns. Here $h_\pm$ are $2\times 2$ matrices
\begin{equation}\label{h4}
h_{s}=\begin{pmatrix}
-ivs\partial_x-\tilde\mu& \Delta(x) \\
\Delta(x) & ivs\partial_x+\tilde\mu
\end{pmatrix},
\end{equation}
and $s=\pm$. It is easy to show that for each allowed value of $k_y$ and each $s$ the above Hamiltonian (\ref{h4}) supports an exact Jackiw-Rossi zero mode \cite{rossi1} with a wave function 
\begin{equation}\label{h5}
\psi_s(x)=A\begin{pmatrix} i\\ s\end{pmatrix}
\exp\left\{-{1\over v}\int_0^xdx'[\Delta(x')-is\tilde\mu]\right\}
\end{equation}
exponentially localized near the junction. 

We shall demonstrate below by an exact numerical diagonalization of a 3D lattice model that the Majorana flat bands remain robust beyond the minimal low energy surface theory. Their  stability however can be deduced from the surface theory alone and relies on the time reversal symmetry $\cT$ and the bulk-boundary correspondence present in the Weyl semimetal.  The most general surface Hamiltonian consistent with these requirements is of the form $h(\bk)=(\bv_{k_y}\cdot\bs) k_x-\eta_{k_y}$ for $|k_y|<K$ where both the velocity vector and $\eta_{k_y}$ are even functions of $k_y$. Time reversal symmetry further permits a term  $(\bu_{k_y}\cdot\bs) k_y$ but this is not consistent with $h(\bk)$ describing the surface state of a Weyl semimetal \cite{remark1}.
For each $k_y$ one can now perform an SU(2) rotation in spin space to bring $h(\bk)$ to the form indicated in Eq.\ (\ref{h2}). Because the pairing term in the BdG Hamiltonian (\ref{h3})  is not affected by this transformation the calculation demonstrating the zero modes proceeds as before, with the result (\ref{h5}) modified in two minor ways: $v$ and $\tilde\mu$ may  now depend on $k_y$ and the spinor structure reflects the SU(2) rotation. The exact zero modes persist for all allowed values of $k_y$ in this more general case. 

It is easy to see that breaking $\cT$ lifts the zero modes: for instance interpolating between $-\Delta_0$ and $+\Delta_0$ through complex values of $\Delta(x)$ moves MZMs to finite energies. Flat Majorana bands are thus protected by $\cT$ and by translation symmetry that is necessary to establish the underlying momentum space Weyl structure.  We argue below that our results are in addition robust against weak nonmagnetic disorder.

\begin{figure*}[t]
\includegraphics[width = 16.5cm]{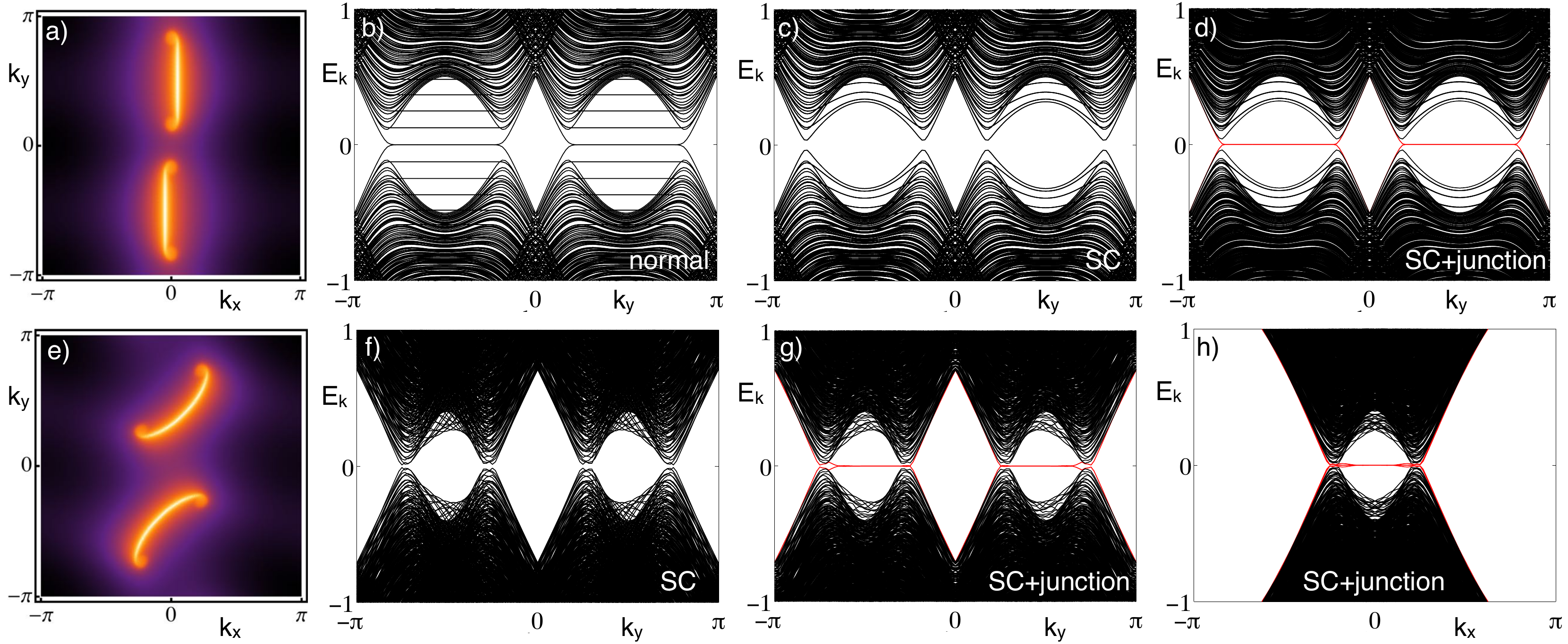}
\caption{Tight binding simulations of the SC/Weyl proximity effect. Panels a) and e) show the surface spectral function $A(\bk,\omega)$ for the tight binding Hamiltonian $h_{\rm latt}+\delta h_{\rm latt}$ and $\omega=0.15$. In all panels we use $m=0.5$, $\lambda=1$,, $L_x=50$ and $L_z=40$; while $(\mu,\epsilon)=(0,0)$ and $(0.1,1.0)$ for top and bottom row respectively.  Panel b) shows the normal state spectrum with flat bands representing the surface states. The effect of the SC proximity effect with $\Delta_0=0.5$ is indicated in panel c) while panel d) shows the effect of two parallel equidistant $\pi$ junctions (protected zero modes indicated in red). The bottom row displays our results for rotated arcs; panel f) the uniform SC surface state, panels g) and h) two parallel  $\pi$ junctions along $y$ and $x$ axes, respectively.  The small gap at Weyl points in panels c) and f) is a finite size effect -- the gap closes as $L_z\to \infty$.
}\label{fig2}
\end{figure*}

To ascertain the validity and robustness of the above analytic results we now study the problem using a lattice model for the Weyl semimetal. We consider electrons in a simple cubic lattice with two orbitals per site and the following minimal Bloch Hamiltonian, 
\begin{equation}\label{h6}
h_{\rm latt}(\bk)=\lambda\sum_{\alpha=x,y,z} s^\alpha\sin{k_\alpha} +\sigma^y s^yM_\bk.
\end{equation}
Here $M_\bk=(m+2-\cos{k_x}-\cos{k_z})$ and $\sigma^\alpha$ are Pauli matrices acting in the orbital space. The Hamiltonian (\ref{h6}) is inspired by Ref.\ \cite{ran1} and adapted to respect $\cT$ (generated here by $is^y\cK$ with $\cK$ the complex conjugation). It has a simple phase diagram.   For $m>\lambda$ (taking $\lambda$ positive) it describes a trivial insulator. At $m=\lambda$ two Dirac points appear at $\bk=(0,\pm\pi/2,0)$ which then split into two pairs of Weyl nodes positioned along the $k_y$ axis. These persist  as long as $|m|<\lambda$.  In this phase surfaces parallel to the $y$ crystal axis exhibit two Fermi arcs terminating at the surface projections of the bulk Weyl nodes illustrated in Fig.\ \ref{fig2}.  

We performed extensive numerical computations based on $h_{\rm latt}(\bk)$ defined above focusing on the slab geometry with surfaces parallel to the $x$-$y$ plane and thickness of $L_z$ sites. The proximity effect is studied using the BdG Hamiltonian (\ref{h1}) with $h_0$  replaced by $h_{\rm latt}$ and $\Delta$ taken to be non-zero in the surface layers only \cite{remark2}. Top row in Fig.\ \ref{fig2} summarizes our results. The normal state shows gapless Fermi arcs (Fig.\ \ref{fig2}a,b) while  uniform SC order is seen to gap out the Fermi arc states except in the vicinity of the Weyl points where they merge into the gapless bulk continuum (Fig.\ \ref{fig2}c). Presence of the $\pi$ junctions (which we define parallel to the $y$ crystal axis) generates perfectly flat bands at zero energy between the projected Weyl nodes, Fig.\ \ref{fig2}d. We note that due to the periodic boundary conditions adopted in both $x$ and $y$ directions our numerics by necessity implement two parallel $\pi$ junctions  which results in twice the number of flat bands compared to a single junction. We also find that, remarkably, the bands remain completely flat even when the system size $L_x$ is small along the direction perpendicular to the junctions (in which case one would naively expect a large overlap between the bound state wavefunctions resulting in significant energy splitting).  This property can be understood by noting that wavefunctions (\ref{h5}) remain {\em exact} zero modes of the Hamiltonian (\ref{h1},\ref{h2}) for periodic boundary conditions along $x$ as long as $\Delta(x)$ averages to zero over all $x$ and $\tilde\mu=(2\pi v/L_x)n$ with $n$ integer. These conditions are satisfied for two equally spaced junctions and $\mu=0$ used in Fig.\ \ref{fig2}. We checked that violating these conditions indeed leads to zero mode splitting that depends exponentially on the junction distance. Importantly, the fact that this detailed property of the simple model (\ref{h2}) is borne out in a more realistic lattice model gives us confidence that the low energy theory provides correct description of the physical surface state of a Weyl semimetal.

The lattice Hamiltonian (\ref{h6}) has high symmetry with Weyl nodes confined to lie on the $k_y$ axis. It is important to verify that the phenomena discussed above are not dependent on such a fine tuned lattice symmetry. To this end we perturb $h_{\rm latt}$ by adding to it 
\begin{equation}\label{h7}
\delta h_{\rm latt}(\bk)=\epsilon\ \sigma^y s^x(1-\cos{k_y}-\cos{k_z}),
\end{equation}
which  respects $\cT$ but breaks the C$_4$ rotation symmetry around the $y$ axis thus allowing Weyl nodes to detach from $k_y$.  For $\epsilon\neq 0$ the arcs rotate and curve as illustrated in Fig.\ \ref{fig2}e. Majorana flat bands however remain robustly present in this low symmetry case (Fig.\ \ref{fig2}g). They now also appear for a junction parallel to the $x$ crystal direction (Fig.\ \ref{fig2}h) because the Weyl points project onto distinct $k_x$ momenta in the boundary BZ. 

We now address the stability of Majorana bands against nonmagnetic disorder that will inevitably be present in any real material. Scaling arguments \cite{burkov1,goswami1,hosur2,syzranov1} and numerical simulations \cite{huang1,sbierski1,hughes1} show that disorder is a strongly irrelevant perturbation in a Weyl semimetal: electron density of states (DOS) at low energies $D(\omega)\sim \omega^2$ remains unchanged up to a critical disorder strength $U_c$ at which point a transition  occurs into a diffusive regime with finite DOS at $\omega=0$. The Fermi arcs likewise remain stable in the weak disorder regime. These theoretical results are confirmed by experimental studies which show clear evidence for Weyl points and Fermi arcs in real materials using momentum resolved probes such as ARPES \cite{hasan1,lv1,yan1,lv2,yang1,hasan2}  and FT-STS \cite{chang1,zheng1},
in agreement with the predictions of momentum space band theory.  In $\cT$-preserving Weyl semimetals one furthermore expects the surface superconducting order to be stable against non-magnetic impurities due to Anderson's theorem \cite{anderson1}. Finally, stability of Majorana flat bands can be argued as follows. In the clean system our calculations show MZMs localized in the vicinity of the junction at each momentum $k$ between the projected Weyl points. Except in the vicinity of the latter these are separated by a gap $\sim\Delta$ from the excited states in the system. Turning on weak random potential of strength $U$ will cause mixing between MZMs at different momenta $k$ as well as with the bulk modes. We expect the former to be a more important effect (except for MZMs in the close vicinity of the Weyl points) because of the low bulk DOS and the assumption of predominantly small momentum scattering.
Importantly, the disorder averaged spectral function of the system must evolve continuously with the disorder strength $U$. For weak disorder, therefore, the Majorana band cannot abruptly disappear; instead disorder will produce a lifetime broadening, giving  the $\delta$-function peak present in the clean spectral function a finite width proportional to $U$. This behavior is indeed observed in numerical simulations of disorder in Majorana flat bands at the edges of 2D topological superconductors \cite{queiroz1}.  We expect the broadened Majorana band to remain observable until its width becomes comparable to the gap $\Delta$ implying a significant range of stability in a weakly disordered sample.

 Majorana flat bands should be directly observable in tunneling spectroscopy of the junction region by a technique developed in the context of other materials \cite{Pill1,Seu1}.
At phase difference $\varphi=\pi$  flat bands will manifest through a distinctive zero-bias peak in the tunneling conductance (with the spectral weight proportional to the length  $K_{\rm flat}$ of the flat portion of the band). The peak will split symmetrically about zero energy as $\varphi$ is tuned away from $\pi$ and will merge into the continuum when $\varphi$ approaches zero.  We note that bulk DOS in a Weyl semimetal $D(\omega)$ is vanishingly small near the $\omega=0$ neutrality point, even in the presence of weak disorder. The prominent zero-bias peak should thus be well visible in a tunneling experiment.
Majorana bands will also have dramatic effect on the current-phase relation $I(\varphi)$ of the junction. Our calculations, to be reported separately \cite{anffany1}, show characteristic step discontinuity in $I(\varphi)$  at $\varphi=\pi$, as the local parity of the junction switches due to the zero mode degeneracy.  The behavior is reminiscent of the current-phase relation predicted to occur in the surface of a topological insulator \cite{ilan1} but the step height in our case will be proportional to $K_{\rm flat}$, directly reflecting the distance between Weyl nodes.
Majorana flat bands also provide a unique opportunity to study interaction effects in Majorana systems because even weak interactions can have profound consequences in a flat-band setting  \cite{potter1,chiu1}. 

We demonstrated the existence and robustness of Majorana flat bands in minimal models with $\cN=4$ Weyl nodes but expect their signatures to survive in materials with larger $\cN$. In this case each Kramers pair of Fermi arcs will produce a Majorana band at zero energy. In the presence of weak disorder there may be some additional broadening due to interband scattering but given that Fermi arcs are quite easy to resolve experimentally \cite{hasan1,lv1,yan1,lv2,yang1,hasan2} even when $\cN=24$ we expect no significant difficulties to arise when $\cN>4$. 

Majorana surface states, including flat bands, have been theoretically predicted to occur in various 2D and 3D topological and nodal superconductors \cite{ReviewQiZhang,RevTanaka,schnyder1,meng1,cho1,fawang1,wong1,you1,hosur1,tanaka1,ikegaya1}. With the exception of high-$T_c$ cuprates, where dispersionless edge states are known to exist (and were reinterpreted recently as Majorana flat bands \cite{potter1}), these have not been observed experimentally because of the general paucity of topological superconductors. Majorana flat bands discussed in this Letter only require ingredients that are currently known to exist -- a $\cT$-preserving Weyl semimetal interfaced with an ordinary superconductor -- and are thus promising candidates for experimental detection. 

\begin{acknowledgments}

The authors are indebted to T. Pereg-Barnea and D.I. Pikulin for discussions, and thank NSERC, CIfAR and Max Planck - UBC Centre for Quantum Materials for support. M.F.\ acknowledges KITP Santa Barbara for hospitality during the initial stages of this project.
\end{acknowledgments}

%...

%%%%%%%%%%%%%%%%%%%%%%%%%%%%%%%%%%%%%%%%%%%%%%%%%%

\begin{thebibliography}{10}
\bibitem{remark1}  For $\bu$ nonzero the surface state exhibits a massless Dirac dispersion which would imply that the Fermi arc can shrink to a point when $\mu$ coincides with the Dirac point.
\bibitem{remark2} Other order parameter configurations (e.g.\ $\Delta$ decaying exponentially into the bulk) were also considered with substantially similar results.
\end{thebibliography}
\end{document}